# Full-dry Flipping Transfer Method for van der Waals Heterostructure

Dohun Kim[1], Soyun Kim[1], Yanni Cho[1], Jaesung Lee[1], Kenji Watanabe[2], Takashi Taniguchi[3], Minkyung Jung[4], Joseph Falson[5], and Youngwook Kim[1*]

[1]*Department of Physics and Chemistry, Daegu Gyeongbuk Institute of Science and Technology (DGIST), Daegu 42988, Republic of Korea*

[2]*Research Center for Functional Materials, National Institute for Materials Science, Tsukuba 305-0044, Japan*

[3]*International Center for Materials Nanoarchitectonics, National Institute for Materials Science, Tsukuba 305-0044, Japan*

[4]*DGIST Research Institute, DGIST, Daegu, Korea*

[5]*Department of Applied Physics and Materials Science, California Institute of Technology, Pasadena, CA 91125, USA*

[*]E-mail: y.kim@dgist.ac.kr

**We present a novel flipping transfer method for van der Waals heterostructures, offering a significant advancement over previous techniques by eliminating the need for polymers and solvents. Here, we utilize commercially available gel film and control its stickiness through oxygen plasma and UV-Ozone treatment, also effectively removing residues from the gel film surface. The cleanliness of the surface is verified through atomic force microscopy. We investigate the quality of our fabricated devices using magnetotransport measurements on graphene/hBN and graphene/α-RuCl$_3$ heterostructures. Remarkably, graphene/hBN devices produced with the flipping method display quality similar to that of fully encapsulated devices. This is evidenced by the presence of a symmetry-broken state at 1 T. Additionally, features of the Hofstadter butterfly were also observed in the second devices. In the case of graphene/α-RuCl$_3$, we observe quantum oscillations with a beating mode and two-channel conduction, consistent with fully encapsulated devices.**

Van der Waals stacking has emerged as a truly disruptive and game-changing technology for device fabrication in the field of two-dimensional electron systems [1, 2]. In conventional epitaxial growth, the substrate imposes the atomic registry of subsequent layers [3, 4]. However, van der Waals stacking eliminates this constraint, allowing for the stacking of two single crystalline layers with vastly different crystal structures, even at a user-selected twist angle[5-17]. As a result, it overcomes the limitations imposed by lattice matching requirements or crystal compatibility in conventional growth methods. This unprecedented capability to combine any layer with another layer offers exceptional diversity without precedent, opening up boundless possibilities in the field.

Novel correlated ground states have recently been observed in van der Waals heterostructures, revealing their potential through electrical transport [5-17] and optical spectroscopy[18-21]. Consequently, there is a pressing need to investigate the local density of states and surface states in order to gain a deeper understanding of these emergent states. However, traditional van der Waals fabrication methods typically involve the full encapsulation of 2D materials with hBN to prevent chemical residues [6, 11-21]. This poses a challenge for conducting scanning tunneling microscopy experiments, which require a conductive surface.

Recent advancements in electrical transport and scanning tunneling microscopy experiments have provided evidence for the effectiveness of flipping transfer methods [22-25]. In one approach, a polymer and tape are utilized to mechanically detach the entire heterostructure [22,23]. Subsequently, the residual polymer is removed through high-temperature annealing in a forming gas environment. Another method involves the use of a combination of poly(vinyl alcohol) layer and PDMS. Once the desired heterostructure is created, it is transferred onto another PDMS substrate, and the poly(vinyl alcohol) layer is dissolved using water [24, 25]. Although both methods facilitate the flipping of the heterostructure, these approaches come with inherent limitations due to their reliance on polymers and the subsequent requirement for polymer removal. Moreover, despite recent demonstrations of full-dry flipping transfer techniques,[26-28] it is important to highlight that certain methods still involve the use of polymers. Additionally, the yield of these methods is not consistently at 100%, often demanding additional heating steps during the transfer process. Importantly, the quality of devices produced by these methods is not adequately demonstrated through electrical transport measurements, which directly assess the device quality and its comparability to fully encapsulated devices.

In this study, we present a novel full-dry flipping transfer method using two gel-paks, where the stickiness of the gel-paks is controlled through oxygen plasma and UV-Ozone treatment. Our approach involves fabricating a graphene/hBN heterostructure and assessing its quality through electrical transport measurements. Remarkably, the performance of our heterostructure is found to be identical to devices manufactured using the conventional pick-up transfer technique [6,7]. Furthermore, we extend our investigation to the electrical transport properties of a graphene/α-RuCl$_3$ heterostructure (top to bottom, after flipping). We observe that the device quality matches that of devices fully encapsulated by hBN [29]. These findings demonstrate our flipping transfer method indeed provides a fully-dry process, preserving an exceptionally clean top metallic surface.

Figure 1 illustrates the flipping transfer technique employed in this study to avoid the use of polymers. To achieve this goal, we utilize the viscoelastic stamp method. Prior to affixing the Gel-pak to the glass substrate, both the glass substrates and one side of the Gel-pak undergo UV-Ozone treatment. This treatment enhances the adhesion at the interface between the glass substrates and the Gel-pak. Afterward, we attach the glass substrate to the Gel-pak, and subsequently, we trim the Gel-pak into two pieces: one measuring 1 × 1 cm and the other 2 × 2 mm in size. The next step involves cleaning the surface residues of the Gel-pak surface.[30,31] These residues are effectively eliminated using two distinct methods as shown in Fig1**a**. For the 1 × 1 cm Gel-pak, we employ oxygen plasma treatment (Gas flow: 50 sccm, Power: 100 W, Pressure during plasma treatment: $3 \times 10^{-1}$ torr, Time: 10 min), a well-established approach that cleans the gel-film surface while maintaining consistent gel-pak stickiness. For the 2 × 2 mm Gel-pak, we utilize UV/Ozone treatment (UV Intensity of 15 mW·cm$^{-2}$ at 185 nm, Time: 5 min), a technique commonly employed in the semiconductor industry for surface treatment. This method not only removes residues but also significantly enhances the stickiness of the gel-pak surface, aligning with our desired condition for the gel-pak film during the heterostructure flipping process.

Consequently, we use the oxygen plasma-treated gel-pak for hBN cleaving, as shown in Fig1**b**-i and Fig1**c**-i, while the UV/Ozone-treated gel-pak is utilized for the target substrate that will receive the 2D heterostructure from the plasma-treated gel-pak. (Note that although the company provides Gel-pak with varying levels of stickiness, their stickiness remains insufficient for transferring 2D materials onto commercially available high stickiness Gel-pak). Before fabricating the flipping transfer, we need to prepare graphene (Fig1**b**-ii and Fig1**c**-ii) and a pre-patterned substrate (Fig2**b**-iii and Fig3**b**-iii), which includes eight Cr/Pt electrodes with a thickness of 2nm and 3nm, respectively.

In contrast to temperature-controlled pick-up and release methods that rely on polymer-based techniques for 2D flake or heterostructure handling [6, 32], our gel-pak based method offers simplicity through direct touch control. For instance, in the case of hBN transfer, hBN is placed on the Gel-pak and then brought into fully contact with the $SiO_2$ substrate, resulting in successful transfer. On the other hand, for picking up a graphene flake, as shown in Fig1**d**, the hBN covers the graphene without making full contact with the $SiO_2$ substrate. The stage is then lifted, facilitating the subsequent transfer process. This process is characteristic of our Fig1**d** procedure. Motorized x, y, and z transfer stages are employed, and the stage temperature is maintained at 80 degrees. Note that our transfer yields are not temperature-dependent. We employ heating primarily to prevent the presence of water molecules during the transfer step. In essence, our method is suitable for use with 2D materials that are sensitive to heat, such as $CrI_3$.

The next step is the flipping process, as shown in Fig1**e**-i. With the prepared UV/Ozone treated Gel-pak, this process is straightforward. The graphene/hBN heterostructure comes into contact with the entire area of the UV/Ozone treated gel-pak, resulting in a flipped heterostructure order of hBN/graphene/Gel-pak. Remarkably, our yield of flipping is 100%. (Note that we conducted flipping experiments on over 20 devices in total, and all of them exhibited successful flipping.) Subsequently, we repeat the simple transfer process to attach the heterostructure to the pre-patterned substrate, as depicted in Fig1**e**-ii, Fig1**f**-i, and Fig1**f**-ii. Graphene becomes the topmost layer. However, further steps are required to connect the pre-patterned electrodes to graphene. Typically, e-beam lithography is employed for this purpose, but in our approach, e-beam resist is not required. Instead, we adapt the 2D transfer process once again to successfully connect the pre-patterned electrodes to graphene. By cleaving graphite onto the plasma-treated gel-pak and transferring cleaved graphite onto the graphene, as shown in Fig1**e**-iii and Fig1**f**-iii. To avoid a low successful yield of graphite transfer caused by significant height variation between the electrodes and the $SiO_2$ substrate, we use 5 nm thick pre-deposited electrodes (2 nm Cr / 3 nm Pt) on the $SiO_2$ substrate, whereas graphene-based heterostructures typically employ 80-100 nm thick electrodes. After completing all transfer processes, we tested the surface using atomic force microscopy, as shown in Fig1**f**-iv (brownish image), and observed no chemical residues on top of the graphene surface. This outcome confirms the achievement of a flipped heterostructure with an exceptionally clean surface and the desired electrical connection, without the necessity of using polymer or annealing steps.

Figure 2**a** illustrates $R_{xx}$ as a function of filling factor at $B = 1$ T, 3 T, and 5 T for Device 1, with our emphasis placed on the electron side. The electron side, characterized by a high mobility of up to 200,000 cm2/Vs, is evident in the positive filling factor region, highlighting superior device quality. However, the hole side also shows a quantitatively similar trend. At 1 T, we observed zero resistance at ν = 2 and 6, along with a resistance hump at ν = 1, indicating symmetry broken states. This feature becomes more pronounced with increasing magnetic field. A clear resistivity minimum is observed at 3 T, and we finally achieve zero resistance at 5 T for ν = 1. At 5 T, we also observed a symmetry broken feature at the second Landau level (ν = 3, 4, ...), which is also evident in the $\sigma_{xy}$ data in Fig 2**b**. Notably, $\sigma_{xy}$ shows a well-defined plateau at ν = 1 and 4, with a hint of a plateau at ν = 3. These results demonstrate that the device quality is comparable to fully encapsulated devices [6,7].

In addition, we fabricated a graphene device, aligning it with hBN to create a Hofstadter butterfly device [7-10, 14]. The mobility values are 165,000 cm$^2$/Vs for the electron side and 130,000 cm$^2$/Vs for the hole side. Secondary Dirac points at +/- 32 V were observed, as shown in Fig 2**c**. Under the application of a magnetic field, we confirmed the development of quantum Hall states not only at the primary Dirac point at 2 V but also at the secondary Dirac points as shown in Fig 2**d**.

Figure 3**a** presents the magnetotransport results of the graphene/α-RuCl$_3$ heterostructure. In this case, the flipping technique proves more powerful than in a simple graphene structure since α-RuCl$_3$ reacts with acetone, making conventional E-beam lithography unlikely. In contrast, our previous approach involved preparing a bottom electrode on top of hBN and cleaning the surface through contact mode atomic force microscopy, a tedious and time-consuming process. However, by using the flipping technique, we avoid the use of solvents and polymers, resulting in a pristine state for α-RuCl$_3$ and significantly reducing the preparation time. Our transport results demonstrate excellent behavior, showing very clear quantum oscillations from $B = 4.5$ T and at $T = 1.5$ K. Notably, a beating mode arising from the spin-split Fermi surface is visible around $B = 8.8$ T, a characteristic feature exclusive to clean devices, this indicates our heterostructure indicate clean interface. Consequently, we observe two distinct FFT peaks, indicated by green and orange triangles in Fig 3**b**. The two FFT peak values (367.7 T, 372.8 T)

correspond to densities of $n = 1.78 \times 10^{13}$ cm$^{-2}$ and $1.80 \times 10^{13}$ cm$^{-2}$, respectively, consistent with previous studies [29,33-39], which observed both the total density and the splitting of the two FFT peak values. [29,37].

Due to α-RuCl$_3$ transferring a significant number of charge carriers to graphene, these heavily doped carriers originate from a hole-doped spin-polarized Fermi pocket at the K point of graphene. Additionally, electron carriers from the Γ point of α-RuCl$_3$ exist in this heterostructure, as revealed by Hall measurements in Fig 3**c** and 3**d** at $T = 1.5$ K. Unlike conventional graphene, it exhibits non-linear behavior. The red solid line represents a linear fit of the original Hall data (solid black line), with the inset showing the residual resistance of the linear fit data. These characteristics indicate the presence of multiple conduction channels in our heterostructure. We used a two-conduction channel model to fit our data in yellow curve of Fig 3**d**, and the electron density at the Γ point is determined to be $3.18 \times 10^{13}$ cm$^{-2}$, which is consistent with our previous study. These results demonstrate that the device manufactured using the flipping transfer method is indeed high quality. Note that the degree of non-linearity in the $R_{xy}$ data does not necessarily indicate the presence or absence of charge transfer. In the case of two-channel conduction, the Hall curve can exhibit significant changes influenced by both carrier densities and mobilities in the two channels.

In conclusion, we have successfully demonstrated a flipping transfer method for van der Waals heterostructures using commercial gel film. Through careful control of the gel film's stickiness via O$_2$ plasma and UV-Ozone treatment, we achieved the efficient transfer of heterostructures from O$_2$ plasma treated gel film to UV-Ozone treated film. Importantly, our approach is entirely dry, eliminating the need for solvents, polymers, or annealing steps. The quality of our devices was evaluated through extensive transport measurements for two examples: graphene/hBN and graphene/α-RuCl$_3$ heterostructures. Both devices exhibited excellent transport behavior comparable to fully encapsulated devices. By eliminating the reliance on wet processes, our flipping transfer technique opens up exciting possibilities for cleaner, and versatile device fabrication in the realm of two-dimensional materials. We believe this method will significantly advance research in van der Waals heterostructures.

**Declaration of competing interest**

The authors declare that they have no known competing financial interests or personal relationships that could have appeared to influence the work reported in this paper.

**Acknowledgment**

This work was supported by the Basic Science Research Program NRF-2020R1C1C1006914 of the Korean Ministry of Science and ICT, and the DGIST-Caltech collaboration research program (23-KUJoint-01). K.W. and T.T. acknowledge support from the JSPS KAKENHI (Grant Numbers 20H00354, 21H05233 and 23H02052) and World Premier International Research Center Initiative (WPI), MEXT, Japan.)

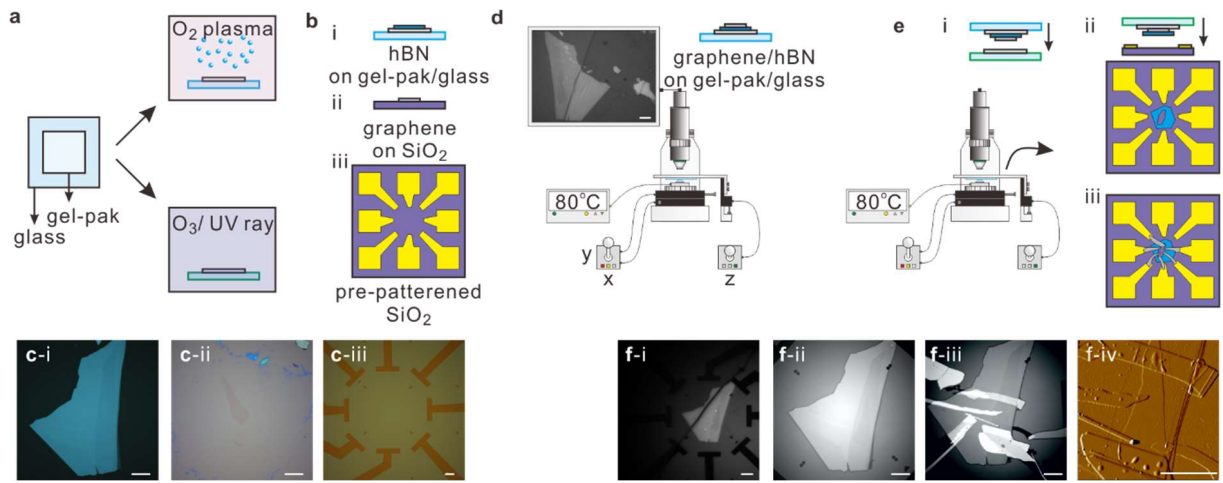

**Figure 1 a**, Two gel-paks on a glass prepared by $O_2$ plasma (up) and UV/Ozone (down) treatment. **b**, Schematic of (**b**-i) hBN on gel-pak/glass, (**b**-ii) graphene on $SiO_2$, and (**b**-iii) pre-patterned $SiO_2$ substrate. **c**, Corresponding optical microscope image of panel **b**. All scale bars are 20 μm. **d,** Schematic of graphene pick up process by using hBN on gel-pak. Microscopy image during pick-up process is displayed with 20 μm scale bar. **e**, Schematic of (**e**-i) the flipping process, (**e**-ii) transfer to patterned substrate, and (**e**-iii) graphite contact process. **f**, Microscopic and atomic force microscopy images (**f**-i) during transfer, (**f**-ii) after transfer, and (**f**-iii, iv). The scale bars of all images are 20 μm.

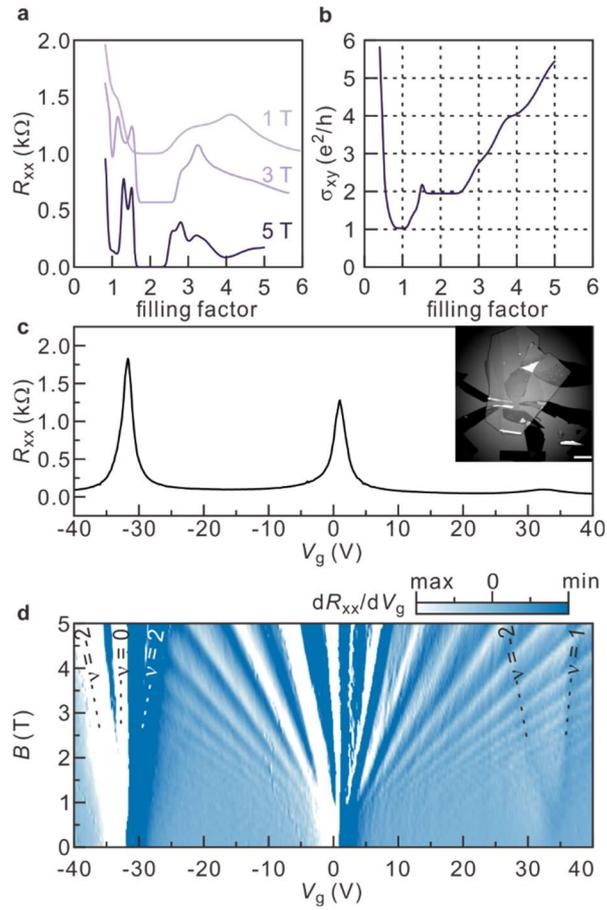

**Figure 2 a**, Longitudinal resistance, $R_{xx}$, as a function of filling factor at $T$ = 1.5 K for Device 1. Different colors indicate different magnetic field. The $R_{xx}$ curves for 3 T and 1 T are vertically offset (0.5 kΩ) for clarity. **b**, $\sigma_{xy}$ curve as a function of filling factor at $B$ = 5 T. **c**, $R_{xx}$ as a function of backgate voltage at $T$ = 1.5 K for Device 2. Resistance peaks at +/- 32V represent mini Dirac points. The inset shows an optical microscope image of Device 2. The scale bar is 20 μm. **d**, Color rendition of $dR_{xx}/dV_g$ as a function of backgate voltage and magnetic field. Dot lines indicate quantum Hall states at mini Dirac points and associated quantum Hall states are written.

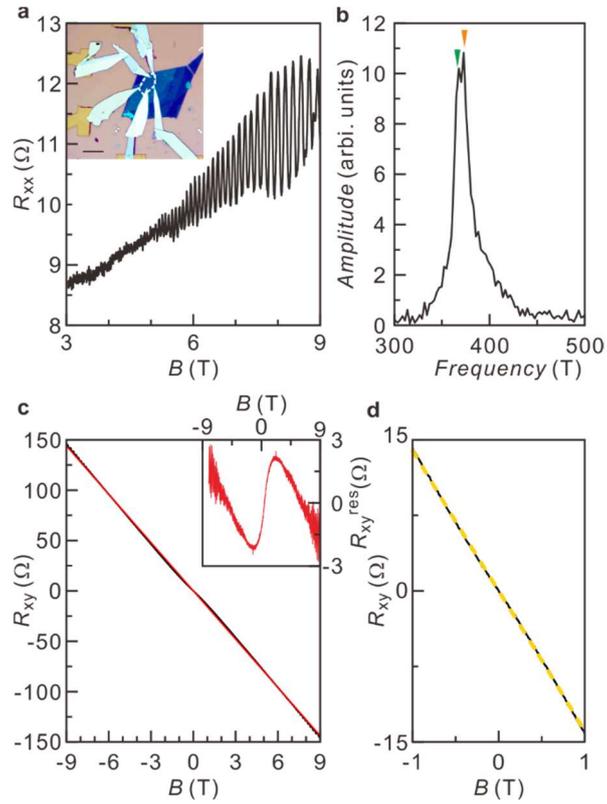

**Figure 3 a**, Longitudinal resistance, $R_{xx}$, as a function of $B$-field at $T = 1.5$ K and $V_g = -70$ V. The inset shows an optical microscope image of the graphene on α-RuCl$_3$. The white dotted line indicates graphene. Large blue flake is α-RuCl$_3$. The scale bar is 20 μm. **b**, FFT spectra of the SdH oscillation in panel **a**. Green and orange triangle show two different frequencies. **c**, $R_{xy}$ as a function of magnetic field at $T = 1.5$ K. Black and red curves indicate original and linear fit data, respectively. Inset shows residual resistance of linear fit. **d**, Same as **c** but with magnified window. Yellow dotted line is two-band model fitted data.